\documentclass[twocolumn]{aastex61}

\usepackage{amsmath}
\usepackage{amsfonts}
\usepackage{amssymb}
\usepackage{float}
\usepackage{graphicx}
\usepackage{calc}
\usepackage[toc,page]{appendix}
\usepackage[section]{placeins}
\usepackage{morefloats}
\usepackage{mathrsfs}
\usepackage{amsfonts}
\usepackage{color}
\usepackage{natbib}

\newcommand{\harm}{{\tt harm}}
\newcommand{\grmonty}{{\tt grmonty}}
\newcommand{\bhlight}{{\tt bhlight}}
\newcommand{\ebhlight}{{\tt ebhlight}}

\newcommand{\mdot}{$\dot{m}$}

\shorttitle{AASTeX 6.1 Template}
\shortauthors{Ryan et al.}

\begin{document}

\title{The Radiative Efficiency and Spectra of Slowly Accreting Black Holes from Two-Temperature GRRMHD Simulations}

\author{Benjamin R. Ryan}
\affil{Department of Astronomy, University of Illinois, 1002 West Green Street, Urbana, IL, 61801}
\author{Sean M. Ressler}
\affiliation{Departments of Astronomy \& Physics, Theoretical Astrophysics Center, University of California, Berkeley, CA 94720}
\author{Joshua C. Dolence}
\affiliation{CCS-2, Los Alamos National Laboratory, P.O. Box 1663, Los Alamos, NM 87545, USA}
\affiliation{Center for Theoretical Astrophysics, Los Alamos National Laboratory, Los Alamos, NM, 87545, USA}
\author{Alexander Tchekhovskoy}
\affiliation{Departments of Astronomy \& Physics, Theoretical Astrophysics Center, University of California, Berkeley, CA 94720}
\affiliation{Lawrence Berkeley National Laboratory, 1 Cyclotron Rd, Berkeley, CA 94720, USA}
\affiliation{Center for Interdisciplinary Exploration \& Research in Astrophysics (CIERA), Physics \& Astronomy, Northwestern University, Evanston, IL 60202, USA}
\author{Charles Gammie}
\affil{Department of Astronomy, University of Illinois, 1002 West Green Street, Urbana, IL, 61801}
\affil{Department of Physics, University of Illinois, 1110 West Green Street, Urbana, IL, 61801}
\author{Eliot Quataert}
\affiliation{Departments of Astronomy \& Physics, Theoretical Astrophysics Center, University of California, Berkeley, CA 94720}

\begin{abstract}
We present axisymmetric numerical simulations of radiatively inefficient accretion flows onto black holes combining general relativity, magnetohydrodynamics, self-consistent electron thermodynamics, and frequency-dependent radiation transport. We investigate a range of accretion rates up to $10^{-5} \dot{M}_{\mathrm{Edd}}$ onto a $10^8 M_{\odot}$ black hole with spin $a_{\star} = 0.5$.  We report on averaged flow thermodynamics as a function of accretion rate. We present the spectra of outgoing radiation and find that it varies strongly with accretion rate, from synchrotron-dominated in the radio at low $\dot{M}$ to inverse Compton-dominated at our highest $\dot{M}$. 
In contrast to canonical analytic models, we find that by $\dot{M} \approx 10^{-5} \dot{M}_{\mathrm{Edd}}$, the flow approaches $\sim 1\%$ radiative efficiency, with much of the radiation due to inverse Compton scattering off Coulomb-heated electrons far from the black hole.
These results have broad implications for modeling of accreting black holes across a large fraction of the accretion rates realized in observed systems.
\end{abstract}

\section{Introduction}

At low mass accretion rates $\dot{m} \equiv \dot{M}/\dot{M}_{\mathrm{Edd}}\footnote{$\dot{M}$ is the accretion rate. $\dot{M}_{\mathrm{Edd}}\equiv 4 \pi G M m_p / \eta \sigma_T c$, where $M$ is the black hole mass and $\eta$ is the nominal efficiency. We adopt $\eta=0.1$; $\dot{M}_{\mathrm{Edd}} = 1.4\times10^{18}(M/M_{\odot})~\mathrm{g/s} = 2.2\times10^{-8}(M/M_{\odot})~M_{\odot}/\mathrm{yr}$.}\lesssim 10^{-3}$, thermally stable accretion onto black holes {is generally believed to form} a geometrically thick, optically thin, radiatively inefficient accretion flow (RIAF, or ADAF\footnote{Advection-Dominated Accretion Flow}; \citealt{Ichimaru1977}, \citealt{NarayanYi1994}, \citealt{YuanNarayan2014}).
Due in part to the two-temperature nature (e.g.\ \citealt{ShapiroLightmanEardley1976}, \citealt{MahadevanQuataert1997}, \citealt{Ressler2015}) of such flows, RIAFs are nearly virial and the liberated gravitational energy is {either} advected across the event horizon or lost through mechanical outflows. Such accretion flows are probably well-represented across the range of astrophysical black hole masses (\citealt{McClintockRemillard2006}, \citealt{Ho2009}).

Analytic and semi-analytic RIAF models have been profitably applied in the study of low-luminosity accretion flows (e.g.\ \citealt{NarayanBarretMcClintock1997}, \citealt{NarayanMahadevanEtAl1998}). However, a limitation of such studies is the reliance on an $\alpha$ viscosity (\citealt{ShakuraSunyaev1973}) to represent angular momentum transport, probably due to magnetohydrodynamic (MHD) turbulence generated by the magnetorotational instability (MRI; \citealt{BalbusHawley1991}). 
Additionally, analytic models typically neglect or approximate general relativity, with potential consequences for interpreting observations as much of a RIAF's outgoing radiation may originate near the black hole ({e.g.\ }\citealt{MoscibrodzkaEtAl2009}).

Global general relativistic numerical simulations have been widely used to study RIAFs driven self-consistently by magnetorotational turbulence (e.g.\ \citealt{KoideShibataKudoh1999}, \citealt{DeVilliersHawleyKrolik2003}, \citealt{McKinneyGammie2004}, \citealt{NarayanEtAl2012}). In the absence of significant mean fields and cooling, such calculations generically recover the hot, nearly Keplerian, nearly axisymmetric (but see \citealt{FragileEtAl2007}) accretion disk {anticipated by analytic models}. {Nonetheless, electron thermodynamics in such calculations has remained a challenge. These flows are collisionless and likely two-temperature (\citealt{Quataert1998}). Historically, constant proton to electron temperature ratios, or other local prescriptions mapping the fluid state to electron temperature (e.g.\ \citealt{MoscibrodzkaEtAl2009}, \citealt{Shcherbakov+2012}, \citealt{Moscibrodzka2014}, \citealt{Chan+2015})} have been employed. Recently, however, \cite{Ressler2015} introduced a method to track numerical dissipation in conservative relativistic MHD schemes, interpret it according to local kinetic plasma studies, and thus separately evolve the electron temperature {(see also \citealt{Sadowski2017} for a similar method). While this provides for physically motivated electron heating, it still assumes a thermal distribution of electrons, whereas these collisionless flows may have a significant population of nonthermal electrons (e.g. \citealt{Kunz2016}, \citealt{Chael2017})}.

Radiative losses are negligible at sufficiently low accretion rates. Towards the Eddington rate, however, radiative processes become important to the dynamics of the flow. Significantly below Eddington, the flow is still optically thin and the electrons are relativistic near the black hole. The dominant energy loss mechanisms are synchrotron emission and Compton upscattering. \cite{Ohsuga2009} first demonstrated that radiation leads to thick/thin disk transitions in numerical models. \cite{FragileMeier2009} found a cooling state inconsistent with either a pure RIAF or a thin disk, and compared {it} to a magnetically-dominated accretion flow in the inner disk. \cite{Moscibrodzka2011} studied accretion rates targeting the supermassive black hole at the center of M87. \cite{Dibi2012} identified $\dot{m} \approx 10^{-7}$ as a critical accretion rate above which radiative losses matter in GRMHD simulations. \cite{Wu2016} targeted the near-Eddington state transition in X-ray binaries in Newtonian MHD with local cooling. Recently, \cite{Sadowski2017} addressed cooling in RIAFs with self-consistent electron heating and a gray M1 radiation closure, while \cite{SadowskiGaspari2017} use a similar model except with constant proton-to-electron temperature ratios to study the {transition to radiatively efficient flows}.

These studies integrate over frequency and adopt a local cooling function or approximate the radiation as a fluid.
In this paper we do not use either of these approximations. Instead, we introduce a scheme that couples a global, albeit axisymmetric, model with electron heating (\citealt{Ressler2015}) for the flow to a Monte Carlo radiation MHD scheme (\citealt{Ryan2015}), yielding a frequency-dependent, full transport solution to the equations of two-temperature relativistic radiation MHD.

We apply this new scheme, \ebhlight{}, to RIAFs across a range of accretion rates. Section \ref{sec:eqns} presents the governing equations. Section \ref{sec:numerics} describes our numerical implementation. Section \ref{sec:results} {contains} our {results}. Section \ref{sec:conclusion} concludes.

\section{Governing Equations}
\label{sec:eqns}

We solve the equations of general relativistic radiation ideal magnetohydrodynamics with full radiation transport. We include a separate electron energy equation (\citealt{Ressler2015}) and electron-photon interactions. {Hereafter, we adopt units such that $c = k_B = 1$ and absorb a factor $\sqrt{4 \pi}$ into definitions of magnetic field strength}.

The radiation and fluid are coupled through exchange of four-momentum. The electron energy density is sourced by numerical dissipation, and electrons and protons exchange energy through Coulomb interactions, as in \cite{Sadowski2017}, allowing transfer of energy between protons and electrons according to the transrelativistic rate of \cite{StepneyGuilbert1983}. Although we track electron and proton temperatures separately, we assume a single four-velocity for the fluid dynamics (\citealt{Ressler2015}).

The dynamical variables in our model are the fluid rest-mass density $\rho_0$, the fluid four-velocity $u^{\mu}$, the fluid internal energy $u$ (equivalently, the fluid pressure $P = (\gamma - 1)u$), the magnetic field three-vector $B^i$,
$\kappa_e \equiv \exp((\gamma_e-1)s_e) = P_e/\rho_0^{\gamma_e}$ ($s_e \equiv$ electron entropy),
  and the radiation specific intensity $I_{\nu}$. We adopt three adiabatic indices: $\gamma_e = 4/3$ for the (relativistic) electrons, $\gamma_p = 5/3$ for the (non-relativistic) protons, and $\gamma = 13/9$ for the total fluid. Although our approximation of three constant $\gamma$ is likely not valid everywhere, previous studies (\citealt{Shiokawa2012}, \citealt{Sadowski2017}) suggest that variable $\gamma$ do not significantly alter conclusions drawn from numerical (GRMHD, GRRMHD) calculations. 

Our full set of governing equations is (written in a coordinate basis):
\begin{align}
\partial_t \left( \sqrt{-g} \rho_0 u^t \right) &= -\partial_i \left( \sqrt{-g} \rho_0 u^i \right), \label{eqn:massConservation}\\
\begin{split}\label{eqn:stressEnergyConservation}
    \partial_t \left( \sqrt{-g} T^t_{~\nu} \right) &={} \partial_i \left( \sqrt{-g} T^i_{~\nu} \right) + \sqrt{-g}T^{\kappa}_{~\lambda} \Gamma^{\lambda}_{~\nu\kappa}\\
& \quad - \sqrt{-g}R^{\mu}_{~\nu;\mu},
\end{split}\\
\partial_t \left( \sqrt{-g} B^i \right) &= \partial_j \left[ \sqrt{-g} \left( b^j u^i - b^i u^j \right) \right], \label{eqn:fluxConservation} \\
\partial_i \left( \sqrt{-g} B^i \right) &= 0, \label{eqn:monopoleConstraint} \\
\frac{dx^{\mu}}{d\lambda} &= k^{\mu}, \label{eqn:photonMotion} \\
\frac{dk^{\mu}}{d\lambda} &= -\Gamma^{\lambda}_{~\mu\nu}k^{\mu}k^{\nu}, \label{eqn:geodesic} \\
\frac{D}{d\lambda}\left(\frac{I_{\nu}}{\nu^3}\right) &= \frac{\eta_{\nu}(T_e)}{\nu^2} -\frac{I_{\nu}\chi_{\nu}(T_e)}{\nu^2}, \label{eqn:radiativeTransfer} \\
\frac{\rho^{\gamma_e}}{\gamma_e - 1} u^{\mu} \partial_{\mu} \kappa_e &= f_e Q_H + Q_C(T_e, T_p) - u^{\nu}R^{\mu}_{~\nu;\mu}, \label{eqn:electronEntropy}
\end{align}
where $D/d\lambda$ is the convective derivative in phase space, the GRMHD stress-energy tensor
\begin{align}
\begin{split}
T^{\mu}_{~\nu} &= \left( \rho_0 + u + P + b^{\lambda}b_{\lambda}\right)u^{\mu}u_{\nu} \\
&\quad + \left(P + \frac{b^{\lambda}b_{\lambda}}{2} \right)g^{\mu}_{~\nu} - b^{\mu}b_{\nu}
\end{split}
\end{align}
with $b^{\mu}$ the magnetic field four-vector (see \citealt{Gammie2003}), and the radiation stress-energy tensor
\begin{align}
R^{\mu}_{~\nu} = \int \frac{d^3 p}{\sqrt{-g}p^t}p^{\mu}p_{\nu} \left(\frac{I_{\nu}}{h^4 \nu^3}\right).
\end{align}
$Q_H$ and $Q_C$ are, respectively, dissipative and Coulomb volumetric heating rates. Temperature dependencies of interaction terms are shown for clarity. Note that $T_e$ is calculated not from $P$ and $\rho_0$ as in \cite{Ryan2015}, but rather from $\rho_0$ and $\kappa_e$ as $T_e = \rho_0^{\gamma_e-1} \kappa_e$. {$T_p = (\gamma_p - 1)(u - u_e)/\rho$ is the proton temperature, only needed for Coulomb coupling.} 
For $T_e = (\gamma_e-1)u_e/\rho_0$, $\Theta_e = m_p T_e / m_e$ where $\Theta_e \equiv$ electron temperature in units of $m_e c^2$.
Note that the radiation four-force  $R^{\mu}_{~\nu;\mu}$ is applied to both the electron and total energy equations; $T^{\mu}_{~\nu}$ incorporates both electrons and protons.

We consider synchrotron emission and absorption. We also include Compton scattering, which for $\Theta_e \gg 1$ and $h \nu \ll k_b T_e$ has a mean amplification factor $\delta E_{\gamma} / E_{\gamma} \approx 16 \Theta_e^2$.

\section{Numerical Method}
\label{sec:numerics}

Our calculations are performed with \ebhlight{}, an extension of \bhlight{} (\citealt{Ryan2015}) that includes the electron heating model of \cite{Ressler2015}. \ebhlight{} solves the equations of GRMHD (Equations \ref{eqn:massConservation} - \ref{eqn:monopoleConstraint}) with the flux-conservative second-order-accurate \harm{} scheme (\citealt{Gammie2003}). The radiative transfer and photon-electron interactions (Equations \ref{eqn:photonMotion} - \ref{eqn:radiativeTransfer}) are evaluated with the Monte Carlo scheme \grmonty{} (\citealt{Dolence2009}; we term radiation samples ``superphotons''). The electron heating (Equation \ref{eqn:electronEntropy}) is evaluated as in \cite{Ressler2015}, with Coulomb heating introduced in a separate explicit second-order step.
We neglect electron and ion conduction, as RIAF simulations have found both to be suppressed by misaligned magnetic fields and temperature gradients (\citealt{Ressler2015}, \citealt{Foucart2017}).
The radiation four-force is evaluated with time-centered fluid quantities, and applied to the total fluid and the electron energy in a first-order operator-split fashion. Emission, absorption, and scattering are treated probabilistically as in \cite{Ryan2015}.

\subsection{Coordinates}

We perform our calculation in horizon-penetrating Modified Kerr-Schild (MKS) coordinates (\citealt{McKinneyGammie2004}). The inner {boundary} is placed inside the event horizon, the outer {boundary at} $r = 200 GM/c^3$. The MKS $h$ parameter is $0.3$. To avoid wasting computational resources advancing many superphotons in the outer region where radiative interactions are relatively unimportant ($\Theta_e \lesssim 1$), we evaluate the radiation sector only inside a smaller outer radius, either $40$ or $100 GM/c^2$, as required to capture at least 95\% of the bolometric luminosity. We employ a spatial resolution $388\times256$ zones.

\subsection{Initial Conditions}

\ebhlight{} is {currently} axisymmetric; the useful time integration window is thus of the order $t\sim1000 GM/c^3$, after which MRI turbulence decays (see \citealt{GuanGammie2007} for details of MRI-driven turbulence in axisymmetry). 
The timescale for viscous electron heating to equilibrate is longer than this beyond $10-15 GM/c^2$.
To address this issue, we initialize our simulation with axisymmetrized data from a 3D {\it nonradiative} GRMHD run with electron heating using the method described in \citealt{Ressler2016}. We consider a low net magnetic flux configuration (i.e.\ SANE rather than MAD; see e.g.\ \citealt{NarayanEtAl2012}). For $\rho_0$, $u$, $\kappa_e$, and $u^i$, axisymmetrization is a straightforward average in $\phi$.
For $B^i$, we construct a vector potential from the 3D data, average that, and then evaluate the axisymmetric field.

The accretion rate is controlled by varying the mass unit conversion from the scale-free GRMHD data. No radiation is present initially; the radiation field equilibrates on the light crossing time. We set the black hole mass to $10^8 M_{\odot}$ (near the turnover of the supermassive black hole mass function; e.g.\ \citealt{Kelly2012}) and dimensionless black hole spin $a_{\star} = 0.5$. 

\subsection{Pathologies}

We employ the drift-frame floors described in \cite{Ressler2016} to repair unphysical {total fluid} densities and energies. Capturing numerical dissipation for electron heating is especially challenging{:} \harm{}-like schemes {can} violate the second law of thermodynamics locally {at the truncation error level}, and in our scheme the electrons may also be {\it cooled} anomalously near large fluid entropy gradients, such as at the funnel wall. See \cite{Ressler2016} for more details. We enforce $\Theta_{e,\mathrm{max}} < 1000$ in the radiation sector, and $T_p/T_e > 0.01$. Additionally, our explicit radiation-fluid coupling may yield negative electron energies. This is difficult to prevent except by increasing superphoton resolution. We monitor such ``supercooling'' events to ensure they are never a significant fraction of the total radiation energy budget. This diagnostic is used to set superphoton resolution, which is related to the cooling time of the flow.

\section{Results}
\label{sec:results}

\begin{deluxetable}{ccccc}
{
\tablehead{
\colhead{$\overline{\dot{m}}$} &
\colhead{$\overline{L}/L_{\mathrm{Edd}}$} &
\colhead{$\overline{\epsilon}$} &
\colhead{$\langle \overline{\Theta_e} \rangle_J$} &
\colhead{$\overline{L_{\mathrm{em}}}/\overline{L_{\mathrm{sc}}}$}
}
\tablecaption{Time-averaged Results\label{tab:results}}
\startdata
    $1.25\times10^{-9}$ & $3.01\times10^{-14}$ & $2.45\times10^{-6}$ & $13.1$ & $1.51\times10^{4}$ \\
    $1.08\times10^{-8}$ & $4.27\times10^{-12}$ & $4.45\times10^{-5}$ & $14.9$ & $1.07\times10^{3}$ \\
    $1.18\times10^{-7}$ & $2.86\times10^{-10}$ & $2.60\times10^{-4}$ & $12.4$ & $1.42\times10^{2}$ \\
    $9.33\times10^{-7}$ & $1.39\times10^{-8}$  & $1.61\times10^{-3}$ & $12.2$ & $1.79\times10^{1}$ \\
    $1.01\times10^{-5}$ & $4.89\times10^{-7}$  & $5.07\times10^{-3}$ & $7.64$ & $1.74$ \\
\enddata
  \tablecomments{{Accretion rate, luminosity, radiative efficiency, emission-weighted electron temperature, and ratio of emission to scattering process{es} for all simulations. Throughout, models are identified by $\overline{\dot{m}}$ rounded to the nearest power of $10$.}}
  }
\end{deluxetable}

We consider the same initial conditions except at five accretion rates: $\dot{m} \sim (10^{-9}, 10^{-8}, 10^{-7},\\ 10^{-6}, 10^{-5})$. Each calculation extends for $1000 GM/c^3$. To {gauge} the importance of cooling, we run these models both with and without radiative cooling. Luminosities from models without cooling are post-processed using \grmonty{} with a $5 GM/c^3$ cadence.

Each superphoton records the $i,j$ indices of the zone of its last interaction; over a time interval $\Delta t$, superphotons captured at the outer radial boundary are used to compute volumetric radiative energy exchange rates in each zone $J_{i,j} \equiv \sum_n -w_n k_{0,n}/(\sqrt{-g}\Delta t \Delta x^1 \Delta x^2 \Delta x^3)$ with the sum taken over the $n$ recorded photons tagged with $i,j$. $J_{\mathrm{em}}$ is that due to emission (with self-absorption subtracted) and $J_{\mathrm{sc}}$ is that due to scattering. As above, heating rates are $Q \equiv du/d\tau$ {due to each process}. Luminosities $L$ are $\int R^1_{~0} \sqrt{-g} dx^2 dx^3$ evaluated at the outer radial radiation boundary. The mass accretion rate $\dot{M} = \int \rho_0 u^1 \sqrt{-g} dx^2 dx^3$ {is} evaluated at the inner radial boundary. Radiative efficiency $\epsilon \equiv L/\dot{M}$.

We begin time averages at the time at which global quantities (\mdot{}, $L$, $\epsilon$) appear relatively steady; time averages (denoted as $\overline{f}$ for a quantity $f$) are always for $600 \leq tc^3/GM \leq 1000$. We also consider weighted spatial averages,
\begin{align}
\langle f \rangle_{\phi} \equiv \frac{\int dx^1dx^2dx^3\sqrt{-g} f \, \phi}{\int dx^1dx^2dx^3\sqrt{-g} \,\phi}.
\end{align}
For simple volume averages inside a radius $r$ denoted $\langle f \rangle_r$, $\phi = 1$ and $r$ sets the upper radial bound of the integrals. $r=10 GM/c^2$ is a natural choice , as it is approximately the radius inside of which viscous equilibrium is achieved.

\begin{figure}
\includegraphics[width=\columnwidth]{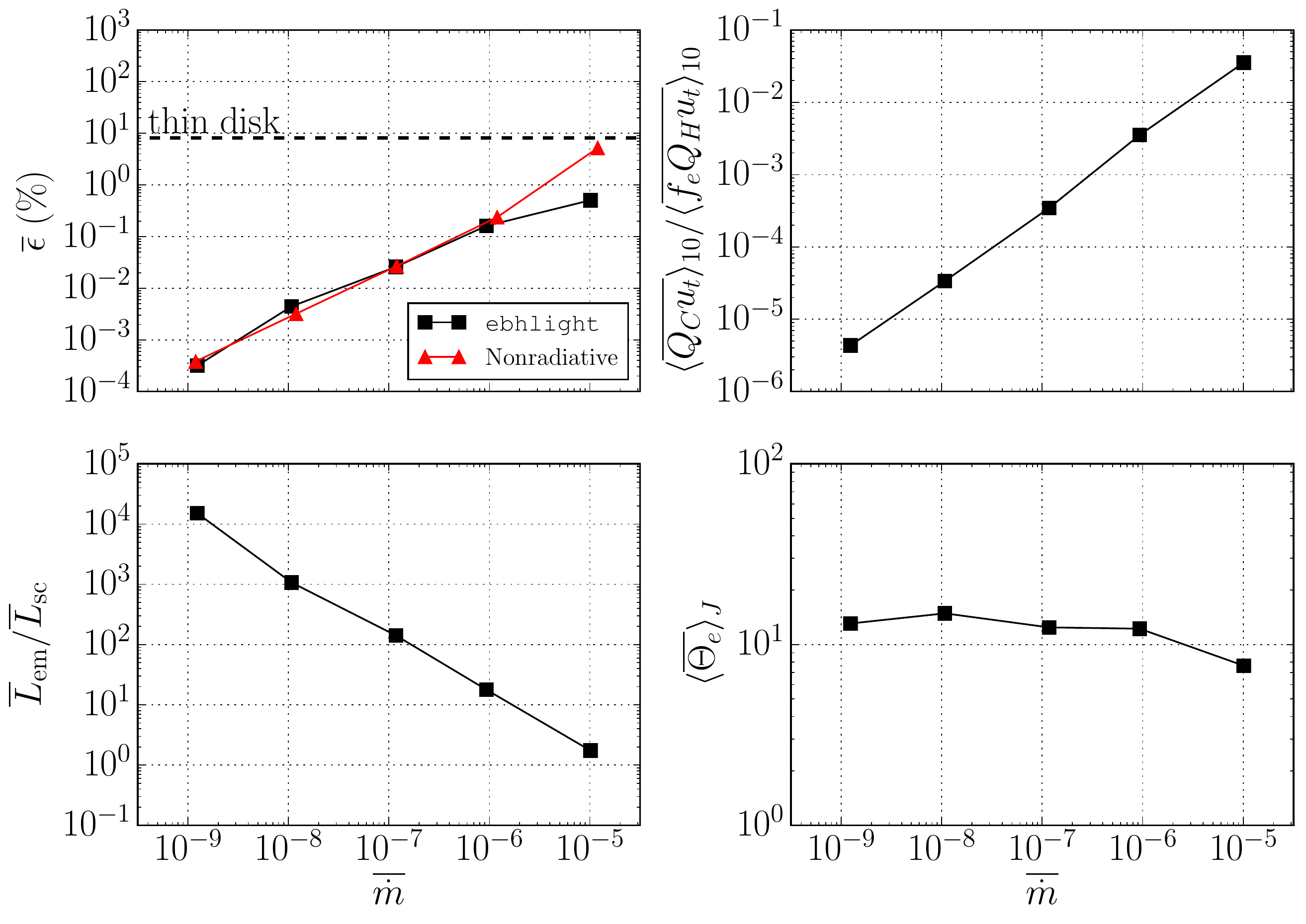}
\caption{Globally averaged quantities. The top left panel shows radiative efficiency $\epsilon$ versus $\overline{\dot{m}}$ for models with and without radiative cooling, along with the thin disk efficiency ($\epsilon = 8.2\%$ for $a_{\star} = 0.5$; \citealt{NovikovThorne}). The top right panel shows the ratio between viscous and Coulomb heating. The bottom right panel shows the emissivity-weighted electron temperature, and the bottom left panel shows the ratio of outgoing radiation due to synchrotron and Compton processes.}
\label{fig:tavgs}
\end{figure}

\begin{figure}
\includegraphics[width=\columnwidth]{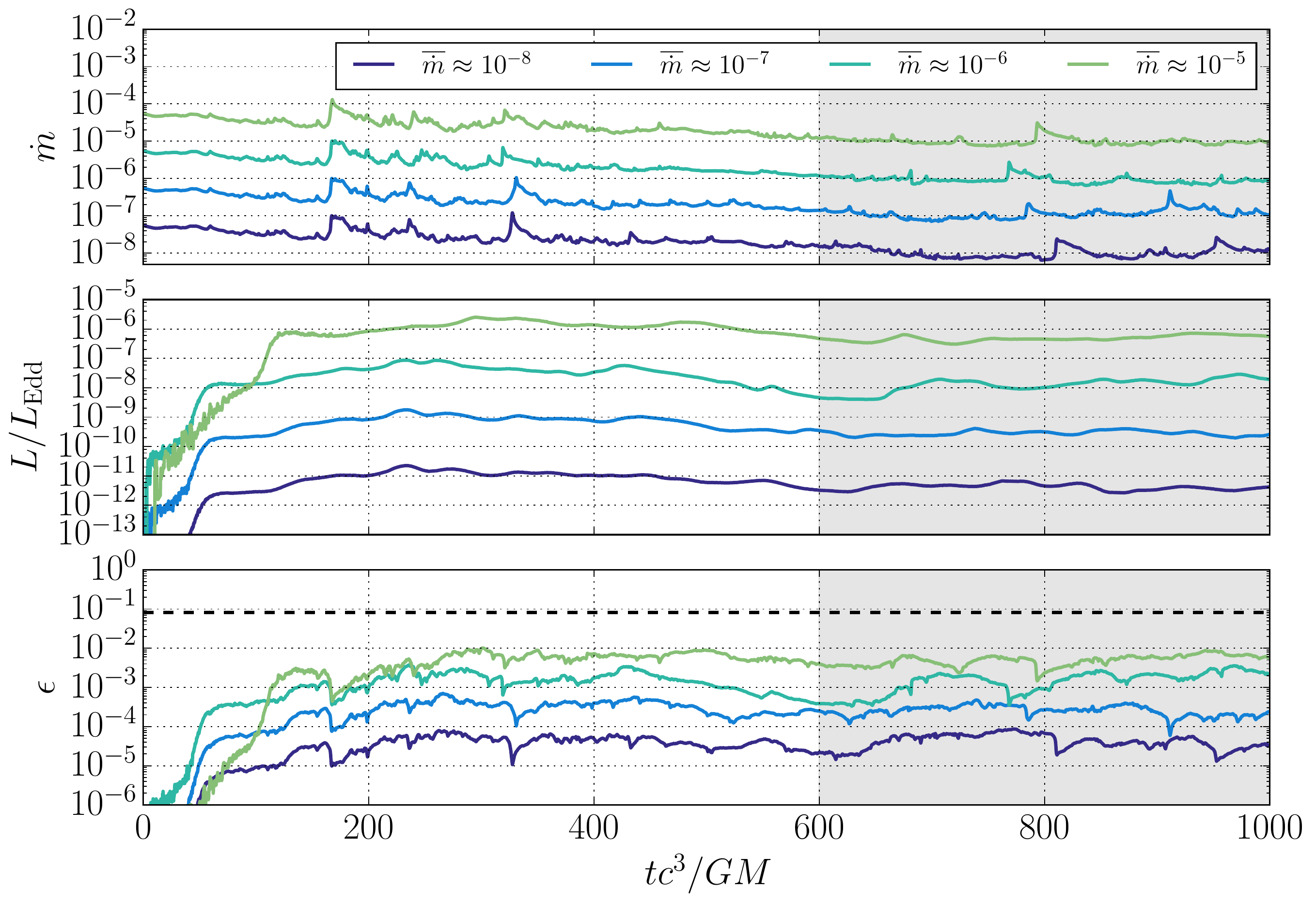}
\caption{Accretion rate, luminosity, and radiative efficiency as a function of time. {Time-averaging window is shown as the shaded region.} {Thin disk efficiency is shown as dashed line in the bottom panel.}}
\label{fig:mdot_lum_eff}
\end{figure}

Figure 1 compares the radiative efficiency $\epsilon$ for radiative and nonradiative models versus $\dot{m}$. Up to $\dot{m} \approx 10^{-6}$, the models are equivalent. At higher $\dot{m}$, however, radiative cooling significantly affects the bolometric luminosity. Therefore, for $\dot{m} \gtrsim 10^{-6}$, self-consistency requires the inclusion of radiative cooling. Note that this value is somewhat higher than the condition $\dot{m} \gtrsim 10^{-7}$ identified by \cite{Dibi2012}, possibly due to differing prescriptions for $T_e$. Additionally, our $\epsilon$ are a factor $\sim 5$ larger at comparable $\dot{m}$ than the $T_p/T_e = 10$, $a_{\star}=0$, 3D models of \cite{SadowskiGaspari2017}. Compton scattering becomes commensurate with synchrotron emission at $\dot{m} \approx 10^{-5}$, and Coulomb heating becomes energetically significant at the $\sim 10\%$ level. The emission-weighted electron temperature $\langle \overline{\Theta_e} \rangle_{J}$ decreases significantly for $\dot{m} \approx 10^{-5}$. 
Electrons inside $r \sim 15 GM/c^2$ achieve thermal equilibrium in our models. These are the radiating electrons for all but the $\dot{m} \approx 10^{-5}$ model, where electrons out to $\sim 30 GM/c^2$ contribute to the luminosity. At $t=1000 GM/c^3$, these electrons are still heating slightly due to Coulomb coupling.
Hereafter we ignore the $\dot{m} \approx 10^{-9}$ model, as flow properties are nearly independent of \mdot{} at such low rates in our model since radiation is negligible. 

Figure \ref{fig:mdot_lum_eff} shows $\dot{m}$, luminosity $L$, and radiative efficiency $\epsilon$ as a function of time. $L$ scales superlinearly with \mdot{} for all simulations reported here {($L \sim \dot{m}^2$, and therefore $\epsilon \sim \dot{m}$, for low \mdot{}, as expected for synchrotron-dominated weak cooling)}{, consistent with the increase in $\epsilon$ with \mdot{} seen in Figure \ref{fig:tavgs}}. Across this range of $\dot{m}$ the flow {transitions} from very radiatively inefficient to a nearly radiatively efficient luminous state.

\begin{figure}
\includegraphics[width=\columnwidth]{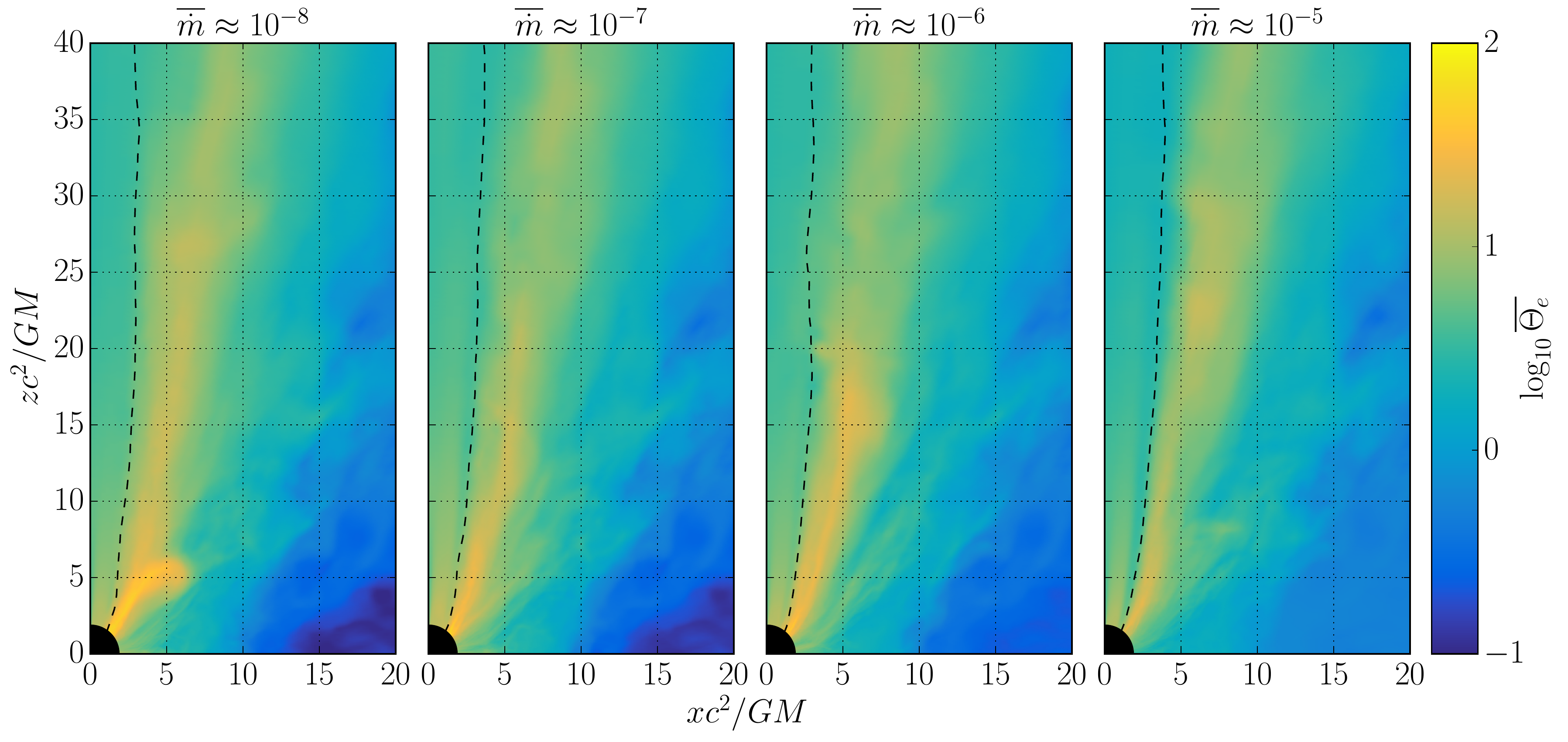}
\caption{Time-averaged electron temperature for all models, also averaged about the midplane. Coulomb collisions heat up the disk at higher \mdot{}
. The dashed line shows the funnel wall, defined as $b^2/\rho = 1$.}
\label{fig:Thetae}
\end{figure}

Figure \ref{fig:Thetae} shows the global structure of {the} electron temperature $\overline{\Theta_e}$ in the accretion flows near the black hole. The electron heating model used here leads to hotter electrons in the more magnetized corona and cooler electrons in the less-magnetized disk midplane (see \citealt{Ressler2015, Ressler2016} for more details). At the highest accretion rates, however, the midplane electrons are significantly hotter (at $r=20 GM/c^2$, in the midplane, $\Theta_e(\dot{m}=10^{-5})/\Theta_e(\dot{m}=10^{-8}) \approx 8$) due to Coulomb heating, and cooling lowers $\Theta_e$ in the inner regions of the flow. 

\begin{figure}
\includegraphics[width=\columnwidth]{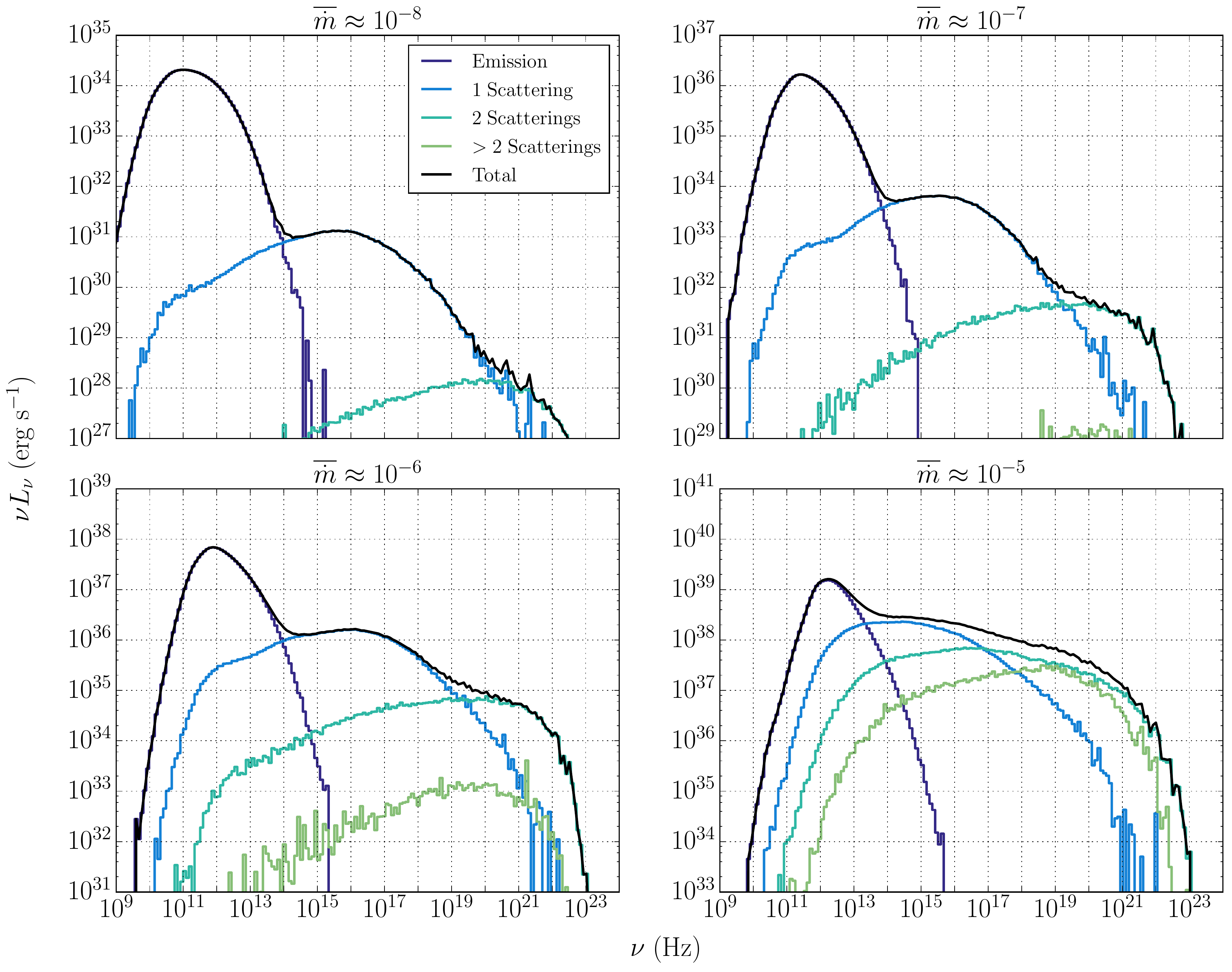}
\caption{Spectral energy distributions for all models. {Both total $\nu L_{\nu}$ and that due to individual interactions (emission, one scattering event, etc.) are shown. The logarithmic interval in $\nu L_{\nu}$ is common to all panels.} At high \mdot{}, multiple Compton scattering events form a high-energy, nearly power-law spectral component}
\label{fig:spectra}
\end{figure}

Figure \ref{fig:spectra} shows the spectra of emergent radiation for an observer nearly in the midplane of the disk. These are evaluated from the same superphotons present in the simulations. At low accretion rates the spectrum is very soft, with distinct Compton bumps, consistent with previous models {where radiation was calculated in post-processing without solving self-consistently for the electron temperature} (e.g.~\citealt{MoscibrodzkaEtAl2009}). As the accretion rate increases, the slope of the high energy tail shifts upwards. 
These trends are consistent with spectral models of 1D RIAFs (e.g.\ \citealt{Esin+1997}, \citealt{Yuan+2004}).

\section{Conclusion}
\label{sec:conclusion}

We have presented general relativistic radiation magnetohydrodynamic simulations of radiatively inefficient black hole accretion flows. We have considered a black hole of mass $10^8 M_{\odot}$ and spin $a_{\star} = 0.5$, and 
accretion rates up to and including those for which radiative cooling is important. In particular, our inclusion of frequency-dependent full radiation transport addresses an important uncertainty in simulations of RIAFs. 

We have found that RIAF models depart from self-consistency at an accretion rate $\dot{m} \approx 10^{-6}${, in the sense that self-consistent calculations with cooling are needed to predict the radiative efficiency and spectrum}. 
By $\dot{m} \approx 10^{-5}$, the cooling of these flows becomes dominated by Compton scattering, rather than emission, and the flow achieves nearly 1\%-level radiative efficiency.

Our results suggest that Coulomb collisions will become as important as viscous heating at an accretion rate of $\dot{m} \approx 10^{-4}$ (extrapolating Figure \ref{fig:tavgs} to somewhat higher \mdot{}).  This is well below what is traditionally assumed in semi-analytic models (for example, \cite{Esin+1997} assume that Coulomb collisions dominate for \mdot{} $\gtrsim$ 0.1).  
This is probably due to the different density and temperature profiles for analytic models and numerical simulations (e.g.\ \citealt{NarayanEtAl2012}).
Future work should study this in 3D simulations and assess the implications of this behavior for observations, including the phenomenology of state transitions in X-ray binaries.

Our study is limited to axisymmetry. To minimize this weakness{,} 
we have used as initial conditions long-duration 3D nonradiative two-temperature GRMHD simulations. 
Nonetheless, we achieve viscous and inflow equilibria only within $r \sim 15 GM/c^2$. This has potential consequences mostly for the $\dot{m} \sim 10^{-5}$ model, for which $\sim 20\%$ of the luminosity is generated beyond $15 GM/c^2$. In this model the electrons at large radius are still heating up; thermal equilibrium would imply a slightly higher radiative efficiency. Should the flow change after viscous equilibration (probably towards reduced proton pressure), the luminosity could be suppressed by $\sim 20\%$, mostly in the high-energy tail of the spectrum.

Our survey is not comprehensive. Black hole mass, spin, accretion disk tilt, and net magnetic flux may all significantly affect these results. We will study these dependencies in future work.

We have directly demonstrated that radiative cooling plays an important role 
in RIAFs. The whole range of accretion rates considered in this work is probably populated by astrophysical sources, and the technique presented here will be valuable in interpreting observations of both stellar mass and supermassive black holes from the mm to the $\gamma$-ray. 

\acknowledgments
It is a pleasure to thank M. Chandra, A. S{\c a}dowski, and J. Stone for useful discussions, as well as the anonymous referee for a very useful report. Work at Los Alamos National Laboratory was done under the auspices of the National Nuclear Security Administration of the US Department of Energy. SMR is supported in part by the NASA Earth and Space Science Fellowship. JD acknowledges support from the Laboratory Directed Research and Development program at Los Alamos National Laboratory. Support for AT was provided by NASA through Einstein Postdoctoral Fellowship grant number PF3-140131 awarded by the Chandra X-ray Center, which is operated by the Smithsonian Astrophysical Observatory for NASA under contract NAS8-03060, and the TAC fellowship, and by NSF through an XSEDE computational time allocation TG-AST100040 on TACC Stampede. This work was made possible by computing time granted by UCB on the Savio cluster. CFG's work was also supported in part by a Romano Professorial Scholar appointment, a Simons Fellowship in Theoretical Physics, and a Visiting Fellowship at All Souls College, Oxford. EQ is supported in part by a Simons Investigator Award from the Simons Foundation and the David and Lucile Packard Foundation. This work was supported in part by NSF grant AST 13-33612. This research used resources provided by the Los Alamos National Laboratory Institutional Computing Program, which is supported by the U.S. Department of Energy National Nuclear Security Administration under Contract No.\ DE-AC52-06NA25396.

\end{document}